\documentclass[prb,showpacs,amsmath,amssymb,amstext,amsfont]{revtex4}
\usepackage{graphicx}
\usepackage{bm}
\usepackage[usenames]{color}
\usepackage{verbatim}
\usepackage{float}
\usepackage{epsf}
\newcommand{\be}{\begin{equation}}
\newcommand{\ee}{\end{equation}}   
\newcommand{\bea}{\begin{eqnarray}}
\newcommand{\eea}{\end{eqnarray}}
\newcommand{\phrl}[1]{Phys.~Rev.~Lett. {\bf #1}}
\newcommand{\phrb}[1]{Phys.~Rev.~B {\bf #1}}

\newcommand{\jpsj}[1]{J.~Phys.~Soc.~Jpn.{\bf #1}}
\newcommand{\jpcm}[1]{J.~Phys.:~Condens.~Matter}
\newcommand{\bib}{\bibitem}
\newcommand{\lb}{\left[}
\newcommand{\rb}{\right]}
\newcommand{\lp}{\left(}
\newcommand{\rp}{\right)}

\newcommand{\br}{\bm{r}}
\newcommand{\bk}{\bm{k}}
\newcommand{\im}{\,\mathrm{Im}\,}
\newcommand{\bn}{\bm{n}}
\def\im{ \mathrm{Im}\, }
\def\re{ \mathrm{Re}\, }

\begin{document}

\title{Quasiparticles near domain walls in hexagonal superconductors}

\author{S. P. Mukherjee and K. V. Samokhin}
\affiliation{Department of Physics, Brock University, St. Catharines, Ontario, Canada L2S 3A1}

\begin{abstract}

We calculate the energy spectrum of quasiparticles trapped by a domain wall separating different time reversal symmetry-breaking ground states in 
a hexagonal superconductor, such as UPt$_3$. The bound state energy is found to be strongly dependent on the gap symmetry, the domain wall orientation, the quasiparticle's direction of semiclassical propagation,
and the phase difference between the domains. We calculate the corresponding density of states and show how one can use its prominent features, in particular, the zero-energy singularity, 
to distinguish between different pairing symmetries.

\end{abstract}

\maketitle

\section{Introduction}
\label{sec:I}

The presence of domain walls (DWs) in a superconductor is a direct evidence of an unconventional nature of the pairing, because the DWs can only appear if there are two or more 
distinct degenerate ground states, which transform one into another by some discrete symmetry operations, \textit{e.g.}, by time reversal. This is possible if the 
superconducting order parameter has more than one component, \textit{i.e.}, corresponds to either a multidimensional irreducible representation (IREP) of the crystal point 
group or to a mixture of different one-dimensional (1D) representations. The chiral $p$-wave state, which is realized, for example, in Sr$_{2}$RuO$_{4}$ (Ref. \onlinecite{Mackenzie}),
is a well-known example of a system in which DWs are believed to play a prominent role. Strong evidence of the superconducting states with broken time reversal symmetry (TRS) has also been reported
in URu$_2$Si$_2$ (Ref. \onlinecite{Schemm}), UPt$_3$ (Ref. \onlinecite{UPt3-review}), SrPtAs (Refs. \onlinecite{SrPtAs-exp} and \onlinecite{Fischer}), and PrOs$_4$Sb$_{12}$ (Refs. \onlinecite{PrOsSb-exp} and \onlinecite{AC07}).
Various TRS-breaking states have been proposed theoretically in Ba$_{1-x}$K$_x$Fe$_2$As$_2$ (Ref. \onlinecite{spis}), doped graphene (Ref. \onlinecite{NLC}), undoped bilayer silicene (Ref. \onlinecite{Liu2}), 
Na$_x$CoO$_2\cdot y$H$_2$O (Ref. \onlinecite{KPHT}), and other materials.  
Superconducting DWs can be created in these systems, \textit{e.g.}, due to the nucleation of the order parameters of opposite chirality in different
parts of an inhomogeneous sample. 

It is well known that a superconducting DW can trap quasiparticles in its vicinity, creating the Andreev bound states (ABS), see, \textit{e.g.}, Ref. \onlinecite{p-wave-ABS}. 
The energy of these states is inside the bulk gap and their very existence can be explained by topological arguments, see Refs. \onlinecite{Volovik} and \onlinecite{Bernevig}.
The ABS contribution to the tunneling density of states (DOS) can be easily separated from that of the bulk quasiparticles and can, therefore, be used to prove the DW presence. Moreover, the ABS spectrum is sensitive to the gap
structure in the bulk of the domains, which allows one to confirm or rule out certain pairing symmetries. 

We focus on the case of a three-dimensional (3D) hexagonal superconductor with the crystallographic point group $D_{6h}$, which describes UPt$_3$.
The quasi-two-dimensional tetragonal case, which is applicable to Sr$_2$RuO$_4$ and the iron-based superconductors, was previously studied in Ref. \onlinecite{Samokhin1}. 
The heavy-fermion superconductor UPt$_3$ has a complicated phase diagram, with two distinct phases (called $A$ and $B$ phases) even in the absence of external magnetic field, see Ref. \onlinecite{UPt3-review} for a review. 
A variety of thermodynamic and transport measurements have revealed an unconventional superconducting state with nodal excitations. 
Although there is still no general consensus on the pairing symmetry in UPt$_3$, the most promising candidate model, which has recently received further 
support from the Josephson interferometry\cite{UPt3-phase} and the polar Kerr effect\cite{UPt3-Kerr} experiments, is based on the two-dimensional (2D) IREP $E_{2u}$ of the point group $D_{6h}$. The corresponding order parameter is real 
in the high-temperature $A$ phase (at 0.45K $<T<$ 0.5K), and complex, \textit{i.e.} TRS-breaking, in the low-temperature $B$ phase (at $T<$ 0.45K). 

Our goal is to study the quasiparticle tunneling features which are uniquely associated with the DWs and can be used to probe the symmetry of the superconducting order parameter, \textit{e.g.}, in the $B$ phase of UPt$_3$.
We analyze the TRS-breaking states corresponding to all IREPs of $D_{6h}$ that support the formation of DWs.
The paper is organized as follows. In Sec. II, we derive a general expression for the ABS energy in the semiclassical (Andreev) approximation. In Sec. III, we calculate the ABS spectrum and the corresponding contribution to the DOS separately
for each of the four possible TRS-breaking states. The summary of our results is presented in Sec. IV. Throughout the paper we use the units in which $\hbar=e=c=1$.

\section{Andreev bound states}
\label{sec:II}

We consider a hexagonal superconductor described by the point group $D_{6h}$, in zero magnetic field. The $z$ axis is along the sixfold symmetry axis and the $xy$ plane coincides with the basal plane. 
The electron band dispersion is assumed to be $\xi(\bk)=(k^2-k_F^2)/2m^*$, where $m^*$ is the effective mass,
with generalization to a more general, \textit{e.g.}, ellipsoidal, case being straightforward. The superconductor is divided into two semi-infinite superconducting domains by a planar DW. 

Since the scale $\xi_d$ of the order parameter variation in
the DW is much greater than the inverse Fermi wavevector, we can use the Andreev approximation,\cite{And64} in which the Bogoliubov quasiparticles propagate along the semiclassical trajectories characterized by the Fermi-surface wavevectors 
$\bk_F=k_{F}(\sin\theta \cos \phi,\sin\theta \sin \phi,\cos\theta)$ ($\theta$ and $\phi$ are the spherical angles, with the polar axis directed along the positive $z$ axis).
In the semiclassical approximation, the gap function is given by a $2\times 2$ spin matrix, which depends on the position $\br$ and the wavevector $\bk_F$: 
$\hat\Delta(\bk_F,\br)=i\hat\sigma_2\psi(\bk_F,\br)$ for singlet pairing, and $\hat\Delta(\bk_F,\br)=i\hat{\bm{\sigma}}\hat\sigma_2\bm{d}(\bk_F,\br)$ for triplet pairing. We consider only 2D IREPs of $D_{6h}$, 
therefore the gap functions can be written in the following form (Ref. \onlinecite{Mineev}): $\psi(\bk_F,\br)=\eta_1(\br)\phi_1(\hat{\bk}_F)+\eta_2(\br)\phi_2(\hat{\bk}_F)$ and 
$\bm{d}(\bk_F,\br)=\eta_1(\br)\bm{\phi}_1(\hat{\bk}_F)+\eta_2(\br)\bm{\phi}_2(\hat{\bk}_F)$, where $\hat{\bk}_F=\bk_F/k_F$,
$\eta_{1,2}$ are the order parameter components, and the basis functions satisfy $\phi_{1,2}(\bk)=\phi_{1,2}(-\bk)$, $\bm{\phi}_{1,2}(\bk)=-\bm{\phi}_{1,2}(-\bk)$. 
The direction of the spin vector $\bm{d}$ is assumed to be fixed along $\hat{\bm{z}}$ by the strong spin-orbit coupling of electrons with the crystal lattice.\cite{z-axis} 

In general, the quasiparticle wave function has four components, corresponding to the Nambu (electron-hole) and spin degrees of freedom, but in the models considered in this paper the spin channels are decoupled and the equations 
are reduced to a two-component form. For each spin projection, the wave function is a product of a rapidly oscillating plane wave $e^{i\bk_F\br}$ and a slowly varying Andreev envelope function $\Psi(\br)$, which satisfies the equation
\be
\lp\begin{array}{cc} -i\bm{v}_F\nabla_{\br} & \Delta_{\bk_F}(\br)\\ \Delta^*_{\bk_F}(\br) & i\bm{v}_F\nabla_{\br} \end{array}\rp \Psi(\br) = E\Psi(\br).
\label{Andreev}
\ee
Here $\bm{v}_F=\bk_F/m^*$ is the Fermi velocity and $\Delta_{\bk_F}(\br)$ is the gap function sensed by the quasiparticles as they propagate along the semiclassical trajectory defined by $\bk_F$: 
$\Delta_{\bk_F}(\br)=\psi(\bk_F,\br)$ for singlet pairing, and $\Delta_{\bk_F}(\br)=d_z(\bk_F,\br)$ for triplet pairing.  

We use the sharp DW model, in which the gap function is described by two different complex constants in the two domains along each semiclassical trajectory:
\begin{equation}
\label{Delta-pm-definition}
  \left.\begin{array}{ll}
        \Delta_{\bk_F}(\br)=\Delta_{+}(\theta, \phi),\quad (\br\cdot\hat{\bf{n}})>0\medskip \\ 
        \Delta_{\bk_F}(\br)=\Delta_{-}(\theta, \phi),\quad (\br\cdot\hat{\bf{n}})<0,
        \end{array}\right.
\end{equation}
where $\hat{\bn}$ is normal to the DW plane. The angular dependence of the gap function is different for different IREPs of the point group, see Sec. \ref{sec:III} below. 
The solution of Eq. (\ref{Delta-pm-definition}) which is exponentially localized near the DW has the form $\Psi_\pm(\br)\sim e^{\mp\kappa_{\pm}(\br\cdot\hat{\bf{n}})}$, where 
$$
  \kappa_{\pm}=\frac{\sqrt{|\Delta_{\pm}|^2-E^2}}{|(\bm{v}_F\cdot\hat{\bf{n}})|}>0.
$$ 
From the continuity of the wave function across the DW, we obtain the following equation for the Andreev bound state (ABS) energy:
\begin{equation}
\label{ABS-energy-eq}
  \frac{E + i(\bm{v}_{F}\cdot\hat{\bn})\kappa_{-}}{E - i(\bm{v}_{F}\cdot\hat{\bn})\kappa_{+}} =\gamma,
\end{equation}
where $\gamma=\Delta_-/\Delta_+=\gamma_R+i\gamma_I$.
The solution of Eq. (\ref{ABS-energy-eq}) is straightforward, see, \textit{e.g.}, Ref. \onlinecite{Samokhin1}, and we find that for each direction of semiclassical propagation satisfying the condition  
\be
\mathrm{sgn}(|\gamma|^2-\gamma_R)\,\mathrm{sgn}(1-\gamma_R)=1,
\label{cond}
\ee
there is exactly one ABS, whose energy is given by
\be
  E_b(\theta,\phi)=|\Delta_{+}(\theta,\phi)|\frac{1}{\sqrt{1+\beta^2(\theta,\phi)}}\ \mathrm{sgn}\lb \beta(\theta,\phi)(\hat{\bm{v}}_{F}\cdot\hat{\bm{n}})\rb,
\label{E_bound}
\ee 
where $\hat{\bm{v}}_F=\bm{v}_F/v_F$ and
\be
\beta(\theta,\phi)= \frac{1-\gamma_{R}}{\gamma_{I}}.
\label{beta-def}
\ee
In all cases studied in this work, we have $|\Delta_+(\theta,\phi)|=|\Delta_-(\theta,\phi)|$, therefore $|\gamma|^2=1$ and the condition (\ref{cond}) is satisfied for every direction. 
It is straightforward to show that the ABS energy (\ref{E_bound}) is inside the bulk gap,
\textit{i.e.}, $|E_b(\theta,\phi)|\leq|\Delta_\pm(\theta,\phi)|$. The dependence of the ABS energy on the direction of semiclassical propagation is not continuous, showing abrupt changes when either $\beta$ or the Fermi
velocity projection on the DW normal change their signs, see Appendix \ref{app: ABS-properties}.

The sharp DW model (\ref{Delta-pm-definition}) can be justified by the following argument. The ABS wave function is exponentially localized on both sides of the DW, with the characteristic scales given by $\kappa_{\pm}^{-1}$. 
The sharp DW approximation is legitimate for those directions of semiclassical propagation for which the DW width $\xi_d$ is smaller than $\kappa_{\pm}^{-1}$. 
This condition is strongly angle-dependent and, in particular, fails for the trajectories corresponding to $(\bm{v}_F\cdot\hat{\bn})\to 0$. However, for such trajectories the Andreev approximation itself is not applicable. 
For most directions of $\bm{k}_F$, one can use the following estimate: $\kappa_{\pm}^{-1}\gtrsim v_F/\Delta_0\sim\xi\sim\xi_d$, where $\Delta_0$ is a characteristic value of the gap and $\xi$ is the superconducting correlation length. 

The quantity of interest is the DOS of the ABS's, which can be measured in tunneling experiments.  
We consider two orientations of the DW, with the normal vector either parallel or perpendicular to the basal plane. In the first case, assuming $\hat{\bm{n}}\parallel\hat{\bm{x}}$, the order parameter depends only on $x$, 
the momentum components parallel to the DW are conserved, and the DOS per unit DW area for both spin projections has the following form:
\begin{equation}
\label{DOS-x-general}
  N_b(E)=\frac{1}{L_yL_z}\sum_{k_{F,y}}\sum_{k_{F,z}}\delta[E-E_b(\bk_F)],
\end{equation}
where $L_y$ and $L_z$ are the system's dimensions in the $yz$ plane. Derivation of this expression is outlined in Appendix \ref{app: DOS-derivation}. Taking the thermodynamic limit and changing the integration variables
to the spherical angles, we obtain:
\be
\label{DOSx}
  N_b(E) = \frac{1}{2}N_Fv_F\int_{0}^{2\pi} d\phi\, |\cos\phi|\int_{0}^{\pi} d\theta\,\sin^2\theta\,\delta[E-E_b(\theta, \phi)],
\ee
where $N_F=m^*k_F/2\pi^2$ is the normal-state DOS in 3D at the Fermi surface per one spin projection. The ABS energy, see Eq. (\ref{E_bound}), has the following form:
\be
  E_b(\theta,\phi)=|\Delta_{+}(\theta,\phi)|\frac{1}{\sqrt{1+\beta^2(\theta,\phi)}}\ \mathrm{sgn}\lb \beta(\theta,\phi)\sin\theta\cos\phi\rb.
\label{Ex_bound}
\ee
In the case of $\hat{\bm{n}}\parallel\hat{\bm{z}}$, the order parameter depends only on $z$, and we obtain the following expressions for the DOS per unit DW area for both spin projections: 
\be
\label{DOSz}
  N_b(E) = \frac{1}{4}N_Fv_F\int_{0}^{2\pi} \ d\phi \int_{0}^{\pi} \ d\theta\,|\sin 2\theta|\,\delta[E-E_b(\theta, \phi)],
\ee
and for the ABS energy:
\be
  E_b(\theta,\phi)=|\Delta_{+}(\theta,\phi)|\frac{1}{\sqrt{1+\beta^2(\theta,\phi)}}\ \mathrm{sgn}\lb \beta(\theta,\phi) \cos \theta\rb.
\label{Ez_bound}
\ee
Due to the electron-hole symmetry, $N_{b}(E)=N_{b}(-E)$, so that below we calculate the DOS only for $E>0$.

\section{Andreev bound states spectra}
\label{sec:III}

The point group $D_{6h}$ has twelve IREPs, six even and six odd, of which eight are 1D and four are 2D. The formation of DWs is possible only for those 
superconducting classes which are degenerate with respect to some discrete symmetry.\cite{VG85,Mineev} Since the 1D IREPs cannot support DWs, we focus 
on the 2D IREPs. We consider only the TRS-breaking chiral states, with the order parameters given by $\bm{\eta}=(\eta_1,\eta_2)\propto\Delta_0(1,\pm i)$.
The momentum dependence of the corresponding gap functions is listed in Table \ref{tab:D6h}. Note that, due to the similarity of the basis functions, our results for the IREPs $E_{1u}$ and $E_{1g}$ are also applicable to 
a tetragonal superconductor with the point group $D_{4h}$.

\begin{table}
\caption{The momentum dependence of the singlet ($\psi$) and triplet ($d_z$) gap functions of the chiral states corresponding to the 2D IREPs of $D_{6h}$, for a strong spin-orbit coupling. The singlet gap functions
correspond to the even IREPs $E_{1g}$ and $E_{2g}$, while the triplet gap functions correspond to the odd IREPs $E_{1u}$ and $E_{2u}$.}
\begin{tabular}{|c|c|}
\hline 
   IREP   & $\psi(\bk)$, $d_z(\bk)$ \\
\hline 
$E_{1u}$  & $k_x\pm ik_y$  \\
\hline 
$E_{1g}$  & $k_{z}(k_x \pm i k_y)$  \\
\hline 
$E_{2u}$ & $k_z(k^{2}_x-k^{2}_y\pm 2ik_xk_y)$ \\
\hline 
$E_{2g}$ & $k^{2}_x-k^{2}_y\pm 2ik_xk_y$  \\
\hline 
\end{tabular}
\label{tab:D6h} 
\end{table}

\subsection{$E_{1u}$ representation}

For the chiral $p$-wave state corresponding to the IREP $E_{1u}$, we find from Table \ref{tab:D6h} the following expressions for the gap functions in the two domains: $\Delta_+=\Delta_0e^{i\chi}(\hat k_{F,x}-i\hat k_{F,y})$ 
and $\Delta_-=\Delta_0(\hat k_{F,x}+i\hat k_{F,y})$. Here $0\leq\chi\leq\pi$ is the Josephson phase difference between the domains, which has to be included in order to satisfy the current conservation across the DW,
see Ref. \onlinecite{Sam12} and also below. Its value depends on the microscopic details of the system, but here we regard it just as an additional phenomenological parameter.
In terms of the spherical angles the gap functions become
\begin{equation}
\left.\begin{array}{l}
      \Delta_{+}(\theta, \phi)=\Delta_0 \sin\theta\, e^{i(\chi-\phi)},\\ 
      \Delta_{-}(\theta, \phi)=\Delta_0 \sin\theta\, e^{i\phi}.
      \end{array}\right.
\end{equation}
It follows from Eq. (\ref{beta-def}) that $\beta=\tan(\phi-\chi/2)$. Below we calculate the ABS energies and the DOS for both orientations of the DW.

\begin{figure}[H]
\vspace{-0.3in}
\includegraphics[width=5in,height=4in, angle=0]{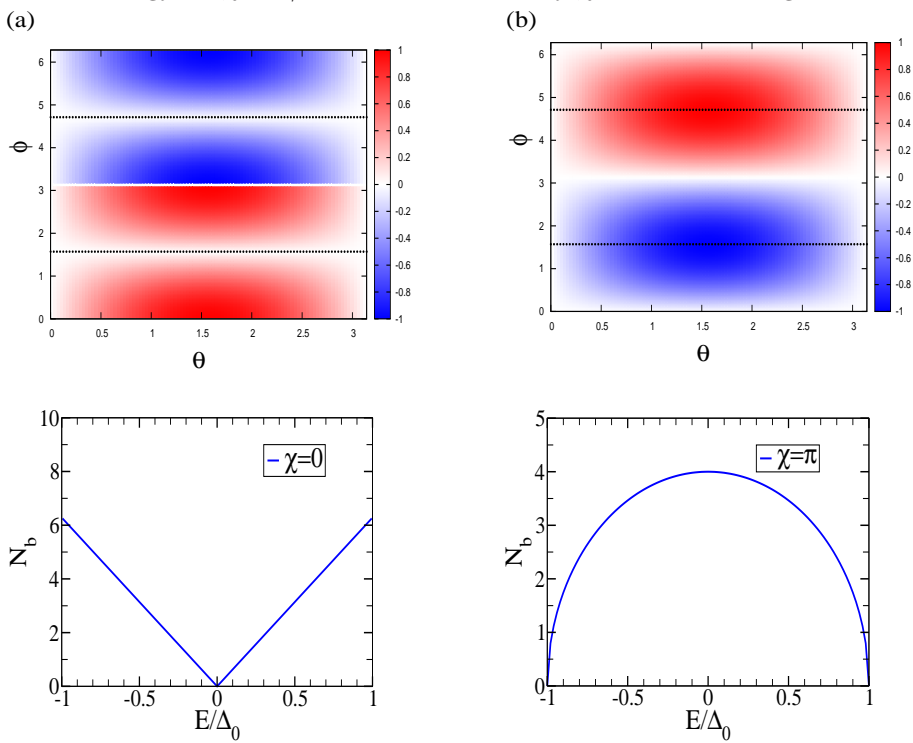}
\caption{(Color online) The ABS energy as a function of the direction of semiclassical propagation (upper panels) and the corresponding DOS (lower panels), 
for $\chi=0$ and $\chi=\pi$, in the case of the chiral $p$-wave state ($E_{1u}$) and $\bm{\hat{n}}||\bm{\hat{x}}$. The grazing trajectories (for which $v_{F,x}=0$)
are shown at $\phi=\pi/2$ and $\phi=3\pi/2$ by horizontal dotted lines.}
\label{fig:E1uC-nx}
\end{figure}

\underline{$\bm{\hat{n}}||\bm{\hat{x}}$.} We obtain from Eq. (\ref{Ex_bound}) the following expression for the ABS energy:
\begin{equation}
\label{Eb-E1u-chi}
  E_b(\theta,\phi)=\Delta_{0}\sin\theta\cos\left(\phi-\frac{\chi}{2}\right)\,\mathrm{sgn}\lb\sin\left(\phi-\frac{\chi}{2}\right)\cos\phi\rb.
\end{equation}
It is shown, for $\chi=0$ and $\chi=\pi$, in the upper panels of Fig. \ref{fig:E1uC-nx}. For general $\chi$, the energy is discontinuous at $\phi-\chi/2=0,\pi$ and also at $\phi=\pi/2,3\pi/2$, see Appendix \ref{app: ABS-properties}.  
The ABS energy has two lines of zeros at $\phi=(\chi\pm\pi)/2$, which correspond to the quasiparticle trajectories with 
$\hat k_{F,x}/\hat k_{F,y}=-\tan(\chi/2)$. These zeros in the ABS dispersion have a topological origin, see Sec. \ref{sec: topology} below. There are also two point zeros in the ABS energy at the poles of the Fermi surface, i.e. at $\theta=0$ and $\pi$.
However, for these directions (corresponding to the ``grazing'' trajectories, parallel to the DW) one has $\hat{\bm{v}}_{F}\cdot\hat{\bm{n}}=0$, and the Andreev calculation resulting in Eq. (\ref{E_bound}) is not applicable. 

The quasiparticle DOS is given by Eq. (\ref{DOSx}) and can be found analytically for $\chi=0$ and $\pi$. Since the calculation is similar in both cases, here we outline it only for $\chi=0$,
when we have
$$
  N_b(E)=N_Fv_F\int_{0}^{\pi} d\theta \sin^2 \theta \int_{0}^{\pi/2} d\phi\cos\phi \lb \delta \lp E-\Delta_{0} \sin\theta \cos \phi \rp + \delta \lp E + \Delta_{0} \sin\theta \cos \phi \rp \rb.
$$
Since we consider only positive energies, the second delta function does not contribute to the integral and we obtain:
\begin{eqnarray*}
  N_b(E) &=& 2N_Fv_F\int_{0}^{\pi/2} d\theta\sin^2\theta \int_{0}^{\pi/2} d\phi\cos\phi\,\delta(E-\Delta_{0} \sin\theta \cos \phi)\\
  &=& \frac{2N_Fv_FE}{\Delta^{2}_{0}} \int_{E/\Delta_{0}}^{1}  \frac{x dx}{\sqrt{1-x^2}\sqrt{x^{2}-(E/\Delta_0)^2}},
\end{eqnarray*}
where $x=\sin\theta$. Evaluating the last integral,\cite{Prudnikov} we finally arrive at the following expression:
\be
\label{DOS-E1u-chi-0}
  N_b(E)=\frac{\pi N_Fv_F}{\Delta_0}\frac{E}{\Delta_0}. 
\ee
In a similar fashion, we obtain:
\be
\label{DOS-E1u-chi-pi}
  N_b(E)= \frac{2N_Fv_F}{\Delta_{0}}\sqrt{1-\frac{E^2}{\Delta^{2}_{0}}}
\ee
for $\chi=\pi$. The DOS curves corresponding to Eqs. (\ref{DOS-E1u-chi-0}) and (\ref{DOS-E1u-chi-pi}), normalized by $N_Fv_F/2\Delta_0$, are shown in the bottom panels of Fig. \ref{fig:E1uC-nx}. We can see that the overall magnitude of the ABS 
contribution to the DOS (per unit area) is of the same order as in the normal state, since $N_Fv_F/\Delta_0\sim N_F\xi\sim N_F\xi_d$.

\underline{$\bm{\hat{n}}||\bm{\hat{z}}$.} Using Eq. (\ref{Ez_bound}) we obtain for the ABS energy:
\begin{equation}
\label{E-E1u-nz}
  E_b(\theta,\phi)=\Delta_{0}|\sin\theta|\cos\left(\phi-\frac{\chi}{2}\right)\,\mathrm{sgn}\lb \sin\left(\phi-\frac{\chi}{2}\right)\cos\theta \rb. 
\end{equation}
It is discontinuous at $\phi-\chi/2=0,\pi$ and also at $\theta=\pi/2$, see Appendix \ref{app: ABS-properties}. 

\begin{figure}[H]
\vspace{-0.3in}
\includegraphics[width=2.5in,height=4in, angle=0]{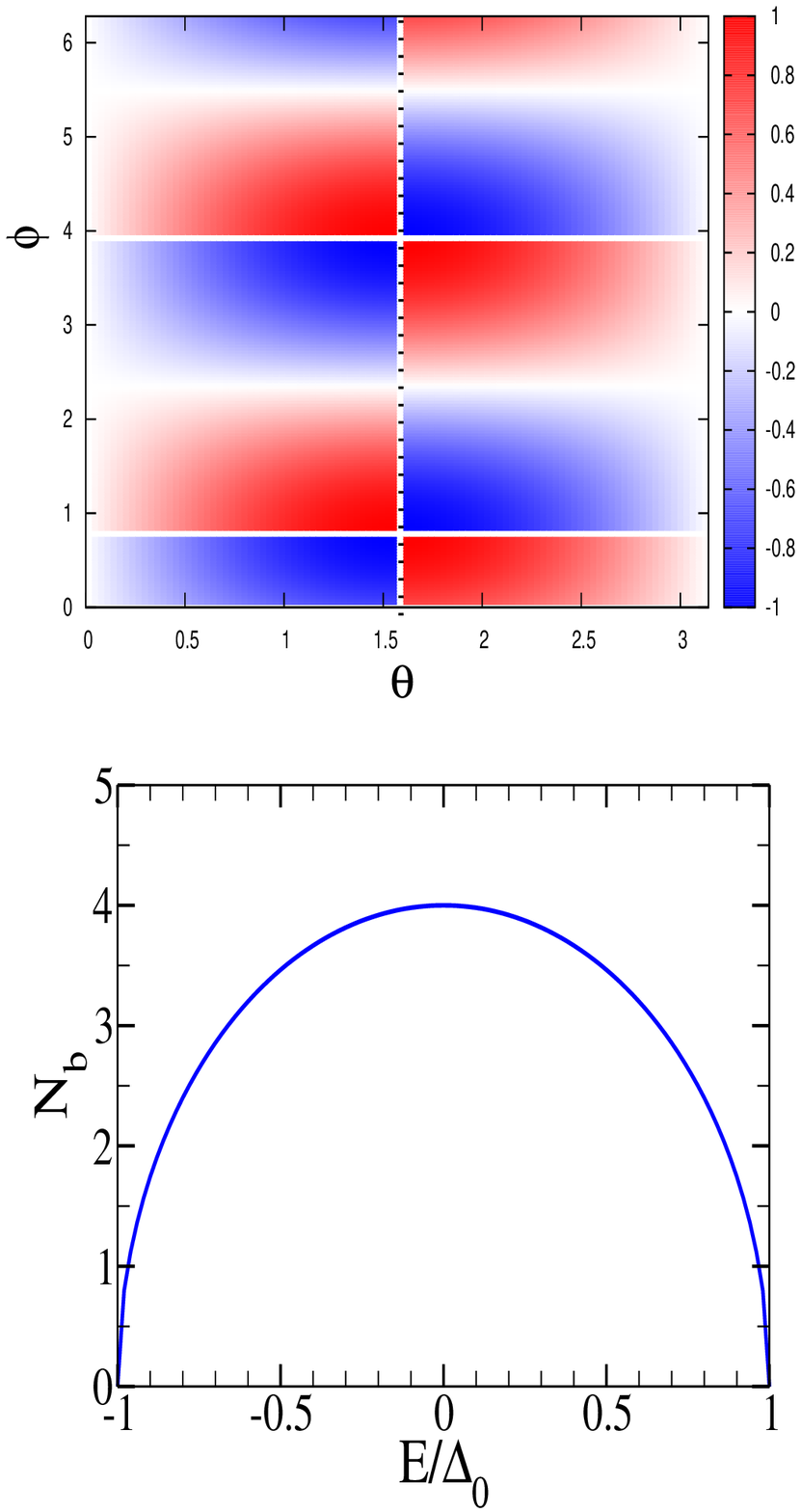}
\caption{(Color online) The ABS energy as a function of the direction of semiclassical propagation for $\chi=\pi/2$ (upper panel) and the corresponding DOS (lower panel), in the case of the chiral $p$-wave state ($E_{1u}$)
and $\bm{\hat{n}}||\bm{\hat{z}}$. The vertical dotted line at $\theta=\pi/2$ corresponds to the grazing trajectory (for which $v_{F,z}=0$).}
\label{fig:E1uC-nz}
\end{figure}

One can show that the current conservation requires that $\chi=\pi/2$ in the lowest order of the Ginzburg-Landau (GL) gradient expansion. The superconducting current can 
be obtained in the standard fashion from the gradient terms in the GL free energy density. Since the order parameter components depend only on $z$, the gradient energy has the form
$F_{grad}=K_4|\nabla_z\bm{\eta}|^2$, in the notations of Ref. \onlinecite{Mineev}. Replacing the gradients by the covariant derivatives, 
$\bm{\nabla}\to\bm{\nabla}+2i\bm{A}$, and varying with respect to the vector potential $\bm{A}$, we obtain for the superconducting current:
$\bm{j}=2K_4\,\mathrm{Im}\,(\bm{\eta}^*\nabla_{z}\bm{\eta})\bm{\hat{z}}$. We use the constant-amplitude approximation for the order parameter: 
$$
  \eta_1(z)=\Delta_0 e^{i\varphi(z)},\quad \eta_2(z)=\Delta_0 e^{i\varphi(z)-i\gamma(z)},
$$
where $\varphi$ is the common (or Josephson) phase of the order parameter components and $\gamma$ is the relative phase, satisfying $\gamma(\pm\infty)=\pm\pi/2$. Then the supercurrent becomes $j_z=2K_4\Delta_0^2(2\nabla_{z}\varphi-\nabla_{z}\gamma)$.
It follows from the current conservation that $j_z$ has a constant value, which is fixed by external sources. Setting $j_z=0$, one obtains $\nabla_z\varphi=\nabla_z\gamma/2$, and, therefore,
\be
  \chi \equiv \varphi(+\infty)-\varphi(-\infty) = \frac{\gamma(+\infty)-\gamma(-\infty)}{2}=\frac{\pi}{2}.
    \label{chi-general}
\ee
It is easy to see that this result holds for the chiral states corresponding to all four 2D IREPs. 
In contrast to the case of $\bm{\hat{n}}||\bm{\hat{x}}$ (Refs. \onlinecite{Sam12} and \onlinecite{Samokhin1}), the Josephson phase difference between the domains for $\bm{\hat{n}}||\bm{\hat{z}}$ takes a universal value $\pi/2$,
\textit{i.e.}, does not depend on the coefficients in the GL expansion. This last conclusion can be invalidated by the inclusion of higher-order gradient terms and going beyond the constant-amplitude approximation. 
However, one can see from the way the angle $\phi$ enters Eq. (\ref{E-E1u-nz}) that the ABS dispersion for different $\chi$ is obtained from that for $\chi=\pi/2$ by simply translating the latter along the $\phi$ 
axis by $\chi/2$. Therefore, the DOS, see Eq. (\ref{DOSz}), does not actually depend on $\chi$. The calculation is similar to the $\bm{\hat{n}}||\bm{\hat{x}}$ case and the final 
result has the following form:
\be
\label{DOS-E1u-nz}
  N_b(E)= \frac{2N_Fv_F}{\Delta_{0}}\sqrt{1-\frac{E^2}{\Delta^{2}_{0}}}.
\ee
The ABS energy for $\chi=\pi/2$ and the DOS for any $\chi$ are shown in Fig. \ref{fig:E1uC-nz}.

\subsection{$E_{1g}$ representation}

For the chiral $d$-wave state corresponding to the IREP $E_{1g}$, we obtain from Table \ref{tab:D6h} the following expressions for the gap functions in the two domains:
$\Delta_+=2\Delta_0e^{i\chi}\hat k_{F,z}(\hat k_{F,x}-i\hat k_{F,y})$ and $\Delta_-=2\Delta_0\hat k_{F,z}(\hat k_{F,x}+i\hat k_{F,y})$. Therefore,
\be
\left.\begin{array}{l}
      \Delta_{+}(\theta, \phi)=\Delta_0 \sin 2\theta\, e^{i(\chi-\phi)},\\ 
      \Delta_{-}(\theta, \phi)=\Delta_0 \sin 2\theta\, e^{i\phi},
      \end{array}\right.
\ee
and $\beta=\tan(\phi-\chi/2)$. 

\begin{figure}[H]
\vspace{-0.3in}
\includegraphics[width=5in,height=4in, angle=0]{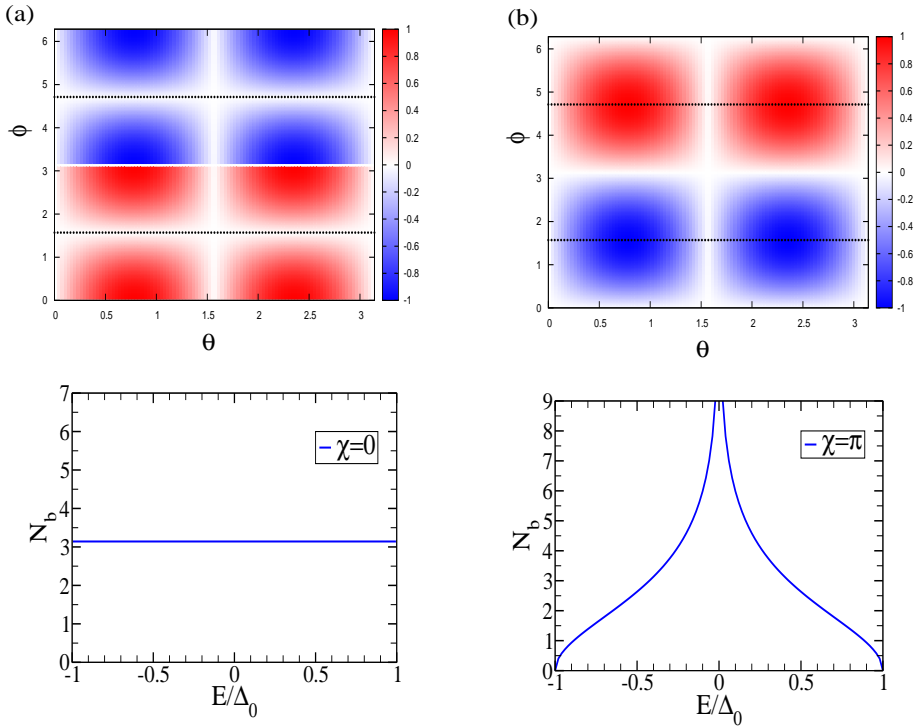}
\caption{(Color online) The ABS energy as a function of the direction of semiclassical propagation (upper panels) and the corresponding DOS (lower panels), 
for $\chi=0$ and $\chi=\pi$, in the case of the chiral $d$-wave state ($E_{1g}$) and $\bm{\hat{n}}||\bm{\hat{x}}$. The grazing trajectories (for which $v_{F,x}=0$)
are shown at $\phi=\pi/2$ and $\phi=3\pi/2$ by horizontal dotted lines.}
\label{fig:E1gC-nx}
\end{figure}

\underline{$\bm{\hat{n}}||\bm{\hat{x}}$.} In this case, Eq. (\ref{Ex_bound}) takes the following form:
\begin{equation}
\label{Eb-E1g-chi}
  E_b(\theta,\phi)=\Delta_{0}|\sin 2\theta|\cos\left(\phi-\frac{\chi}{2}\right)\,\mathrm{sgn}\lb\sin\left(\phi-\frac{\chi}{2}\right)\cos\phi\rb.
\end{equation}
The ABS energy is discontinuous at $\phi-\chi/2=0,\pi$ and also at $\phi=\pi/2,3\pi/2$, see Appendix \ref{app: ABS-properties}. 
It has two lines of zeros at $\phi=(\chi\pm\pi)/2$ and another one in the basal plane, i.e., at $\hat k_{F,z}=0$. The point zeros at $\theta=0$ and $\pi$ correspond to the trajectories parallel to the DW,
for which the Andreev approximation is not applicable. In the upper panels of Fig. \ref{fig:E1gC-nx}, we show the ABS energy for $\chi=0$ and $\pi$. 

The DOS is given by Eq. (\ref{DOSx}). Following the same steps as in the previous subsection, we obtain a constant DOS:
\be
  N_b(E)=\frac{\pi N_Fv_F}{2\Delta_{0}}
\ee
for $\chi=0$, see the bottom panel of Fig. \ref{fig:E1gC-nx}. For $\chi=\pi$, we have from Eq. (\ref{Eb-E1g-chi}): $E_b(\theta,\phi)=\Delta_{0}|\sin 2\theta|\sin\phi$, and Eq. (\ref{DOSx}) can be reduced to the form
$$
  N_b(E)=2N_Fv_F\int_{0}^{\pi/2} d\theta \sin^2\theta \int_{0}^{\pi/2} d\phi\cos\phi\,\delta\lp E-\Delta_{0}\sin2\theta\sin\phi\rp
  =\frac{N_Fv_F}{\Delta_0}\int_{\alpha}^{\pi/2-\alpha} d\theta\tan\theta,
$$	
where $\alpha=(1/2)\arcsin(E/\Delta_0)$. The last integral can be easily evaluated and we arrive at the following final expression for the DOS:
\be
N_b(E)= \frac{N_Fv_F}{\Delta_0}\ln\cot\left(\frac{1}{2}\arcsin\frac{E}{\Delta_0}\right),
\ee
which diverges logarithmically at $E\rightarrow 0$, as shown Fig. \ref{fig:E1gC-nx}. This divergence is nothing but the van Hove singularity due to the saddle points in the ABS dispersion 
at $\theta=\pi/2$ and $\phi=0,\pi$, \textit{i.e.}, for $\hat{\bk}_F$ perpendicular to the DW. 

\begin{figure}[H]
\vspace{-0.3in}
\includegraphics[width=2.5in,height=4in, angle=0]{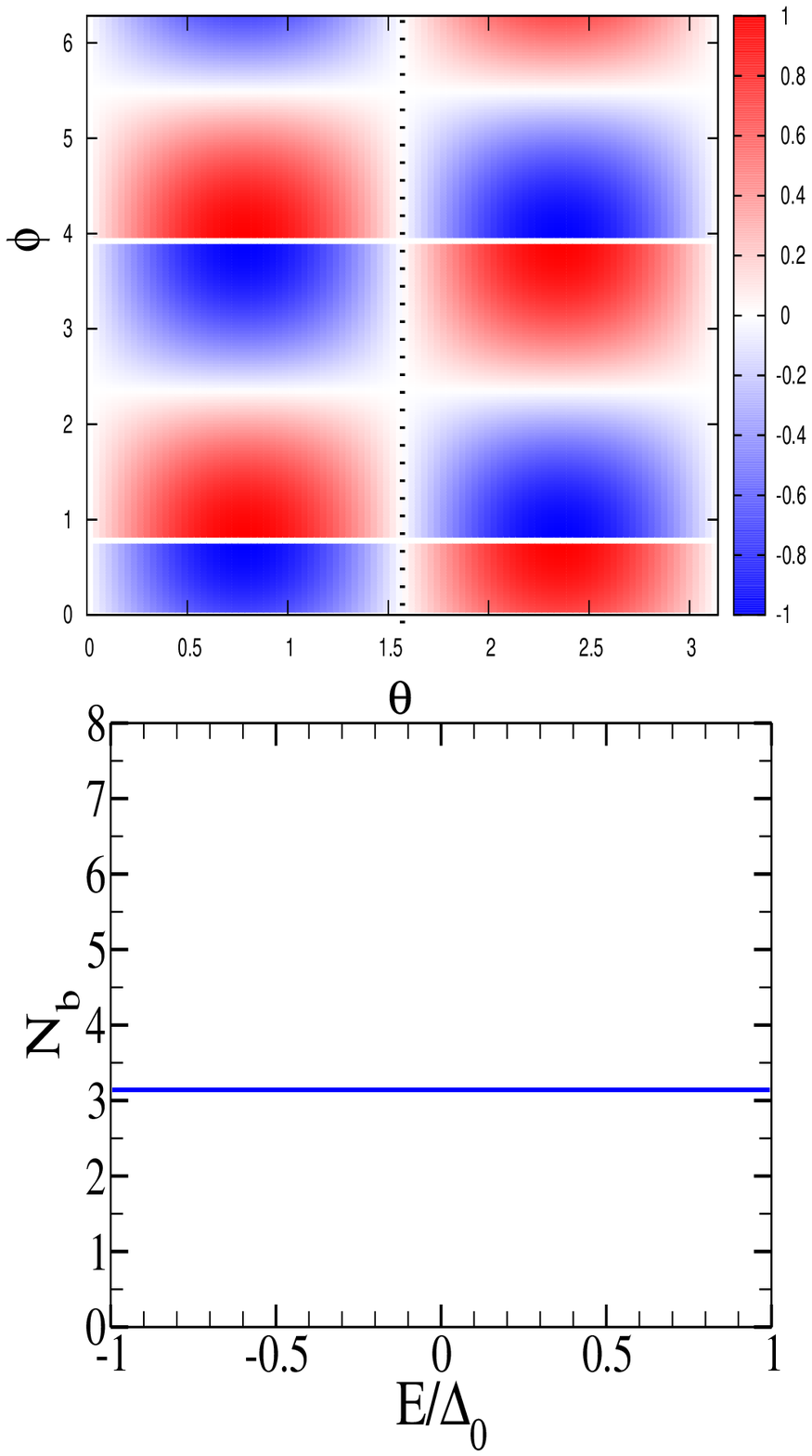}
\caption{(Color online) The ABS energy as a function of the direction of semiclassical propagation for $\chi=\pi/2$ (upper panel) and the corresponding DOS (lower panel), in the case of the chiral $d$-wave state ($E_{1g}$)
and $\bm{\hat{n}}||\bm{\hat{z}}$. The vertical dotted line at $\theta=\pi/2$ corresponds to the grazing trajectory (for which $v_{F,z}=0$).}
\label{fig:E1gC-nz}
\end{figure}

\underline{$\bm{\hat{n}}||\bm{\hat{z}}$.}  We obtain from Eq. (\ref{Ez_bound}):
\begin{equation}
  E_b(\theta,\phi)=\Delta_{0}\sin 2\theta\cos\left(\phi-\frac{\chi}{2}\right)\,\mathrm{sgn}\lb\sin\left(\phi-\frac{\chi}{2}\right)\rb.
\end{equation}
The energy is discontinuous at $\phi-\chi/2=0$ and $\pi$, see Appendix \ref{app: ABS-properties}. 
As for the $E_{1u}$ IREP, the ABS dispersion as a function of $\phi$ simply shifts upon changing $\chi$, therefore the DOS does not depend on $\chi$. A straightforward calculation yields the following result:
\be
  N_b(E)=\frac{\pi N_Fv_F}{2\Delta_{0}}.
\ee
The ABS energy for $\chi=\pi/2$ and the DOS for any $\chi$ are shown in Fig. \ref{fig:E1gC-nz}.

\subsection{$E_{2u}$ representation}
\label{sec: E-2u}

For the chiral $f$-wave state corresponding to the IREP $E_{2u}$, we obtain from Table \ref{tab:D6h} the following expressions for the gap functions in the two domains:
$\Delta_+=\Delta_0e^{i\chi}\hat k_{F,z}(\hat k^2_{F,x}-\hat k^2_{F,y}-2i\hat k_{F,x}\hat k_{F,y})$ and $\Delta_-=\Delta_0\hat k_{F,z}(\hat k^2_{F,x}-\hat k^2_{F,y}+2i\hat k_{F,x}\hat k_{F,y})$. Therefore,
\be
\left.\begin{array}{l}
      \Delta_{+}(\theta, \phi)=\Delta_0 \sin^2\theta\cos\theta\, e^{i(\chi-2\phi)},\\ 
      \Delta_{-}(\theta, \phi)=\Delta_0 \sin^2\theta\cos\theta\, e^{2i\phi},
      \end{array}\right.
\ee
and $\beta=\tan(2\phi-\chi/2)$. 

\begin{figure}[H]
\vspace{-0.3in}
\includegraphics[width=5in,height=4in, angle=0]{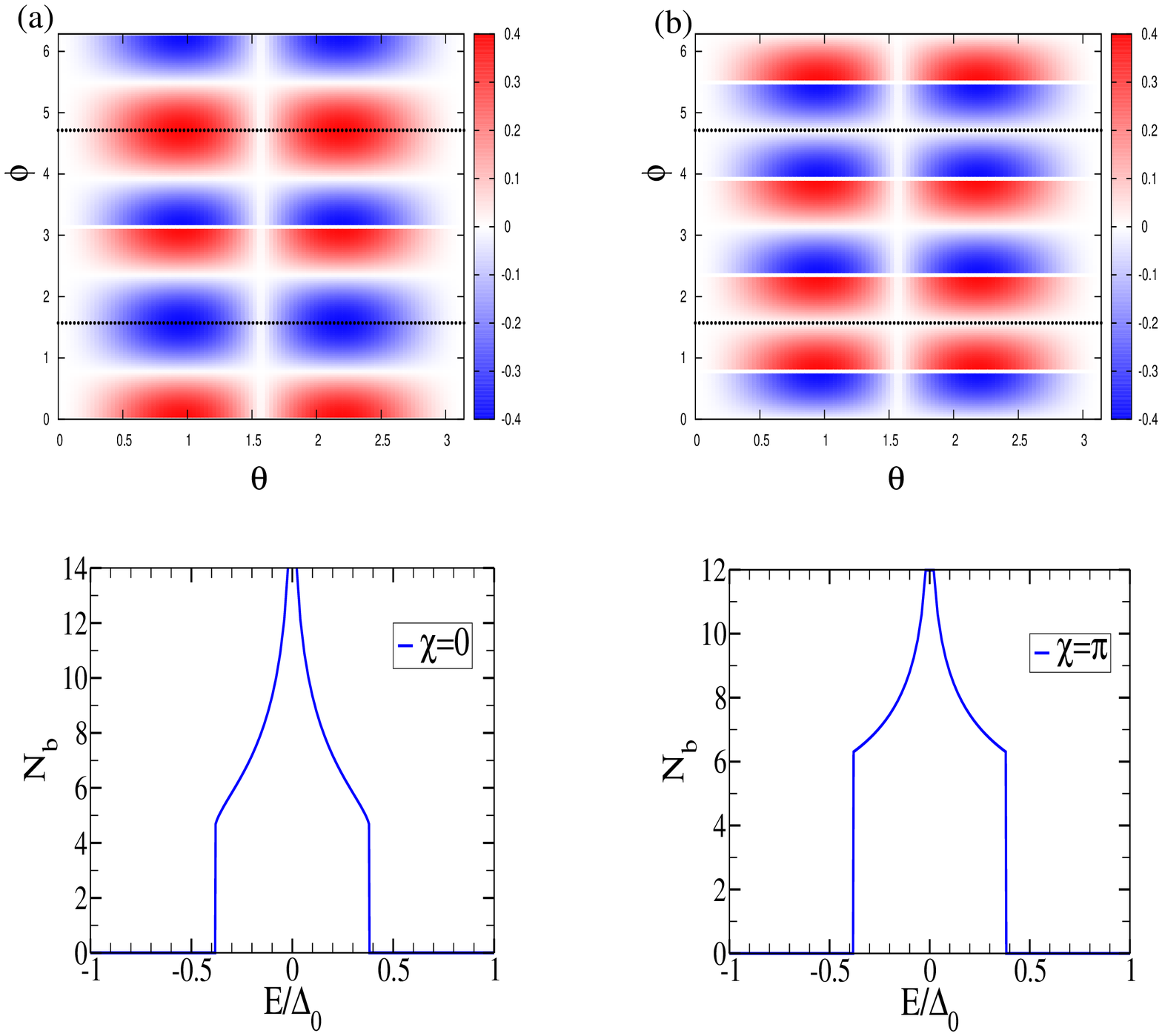}
\caption{The ABS energy as a function of the direction of semiclassical propagation (upper panels) and the corresponding DOS (lower panels), 
for $\chi=0$ and $\chi=\pi$, in the case of the chiral $f$-wave state ($E_{2u}$) and $\bm{\hat{n}}||\bm{\hat{x}}$. The grazing trajectories (for which $v_{F,x}=0$)
are shown at $\phi=\pi/2$ and $\phi=3\pi/2$ by horizontal dotted lines.}
\label{fig:E2uC-nx}
\end{figure}

\underline{$\bm{\hat{n}}||\bm{\hat{x}}$.} In this case, Eq. (\ref{Ex_bound}) takes the following form:
\begin{equation}
\label{Eb-E2u-chi}
  E_b(\theta,\phi)=\Delta_{0}\sin^2\theta|\cos\theta|\cos\left(2\phi-\frac{\chi}{2}\right)\,\mathrm{sgn}\lb\sin\left(2\phi-\frac{\chi}{2}\right)\cos\phi\rb.
\end{equation}
This expression has discontinuities at $2\phi-\chi/2=0,\pi$ and also at $\phi=\pi/2,3\pi/2$, see Appendix \ref{app: ABS-properties}. 
It has four lines of zeros at $\phi=(\chi\pm\pi)/4$ and $\phi=(\chi\pm 3\pi)/4$, and another one at $\hat k_{F,z}=0$.  The isolated second-order point zeros at $\theta=0$ and $\pi$ correspond to the trajectories parallel to the DW,
for which the Andreev approximation is not applicable. 
The ABS energy for $\chi=0$ and $\pi$ is shown in the upper panels of Fig. \ref{fig:E2uC-nx}. 

The DOS for $\chi=0$, see Eq. (\ref{DOSx}), can be reduced to the following form:
\be
\label{DOS-E2u-chi-0}
N_b(E)= \frac{N_Fv_F}{\sqrt{2}\Delta_{0}} \int_{0}^{1} \frac{dx}{\sqrt{x}}\left[\frac{1}{\sqrt{x(1-x^2)-(E/\Delta_0)}}+\frac{1}{\sqrt{x(1-x^2)+(E/\Delta_{0})}}\right] \Theta\left[x(1-x^2)-\frac{E}{\Delta_0}\right],
\ee
where $\Theta(x)$ is the Heaviside step function and $x=\cos\theta$.
Since the function $x(1-x^2)$ attains its maximum at $x=1/\sqrt{3}$, the DOS vanishes at $E>2\Delta_{0}/3\sqrt{3}$. The integral in Eq. (\ref{DOS-E2u-chi-0}) is evaluated numerically.
The logarithmic van Hove singularity in the DOS at $E\to 0$ is due to the saddle points in the ABS dispersion in the basal plane, at $\theta=\pi/2$ and $\cos(2\phi)=0$.

One can easily show that the DOS has a zero-energy singularity at all values of $\chi$. Indeed, we have 
$$
  |\bm{\nabla}E_b|=\Delta_0\sin\theta\sqrt{(3\cos^2\theta-1)^2\cos^2(2\phi-\chi/2)+\sin^2(2\theta)\sin^2(2\phi-\chi/2)},
$$
away from the spectrum discontinuities.
This last expression has the following zeros: (i) $\theta=0,\pi$, whose contribution to the DOS is nonsingular, due to the factor in front of the $\delta$-function in Eq. (\ref{DOSx}); (ii) $\cos^2\theta=1/3$ and $\sin(2\phi-\chi/2)=0$,
which corresponds to a maximum (minimum) of $E_b$; and (iii) $\theta=\pi/2$ and $\cos(2\phi-\chi/2)=0$, which corresponds to the saddle points of $E_b$. It is the saddle points, which are located at the four perpendicular directions in the basal plane 
where the lines of zeros of $E_b$ intersect, that produce the van Hove singularity at $E\to 0$. The DOS for $\chi=0$ and $\chi=\pi$ are shown in Fig.~\ref{fig:E2uC-nx}. 

\begin{figure}[H]
\vspace{-0.3in}
\includegraphics[width=2.5in,height=4in, angle=0]{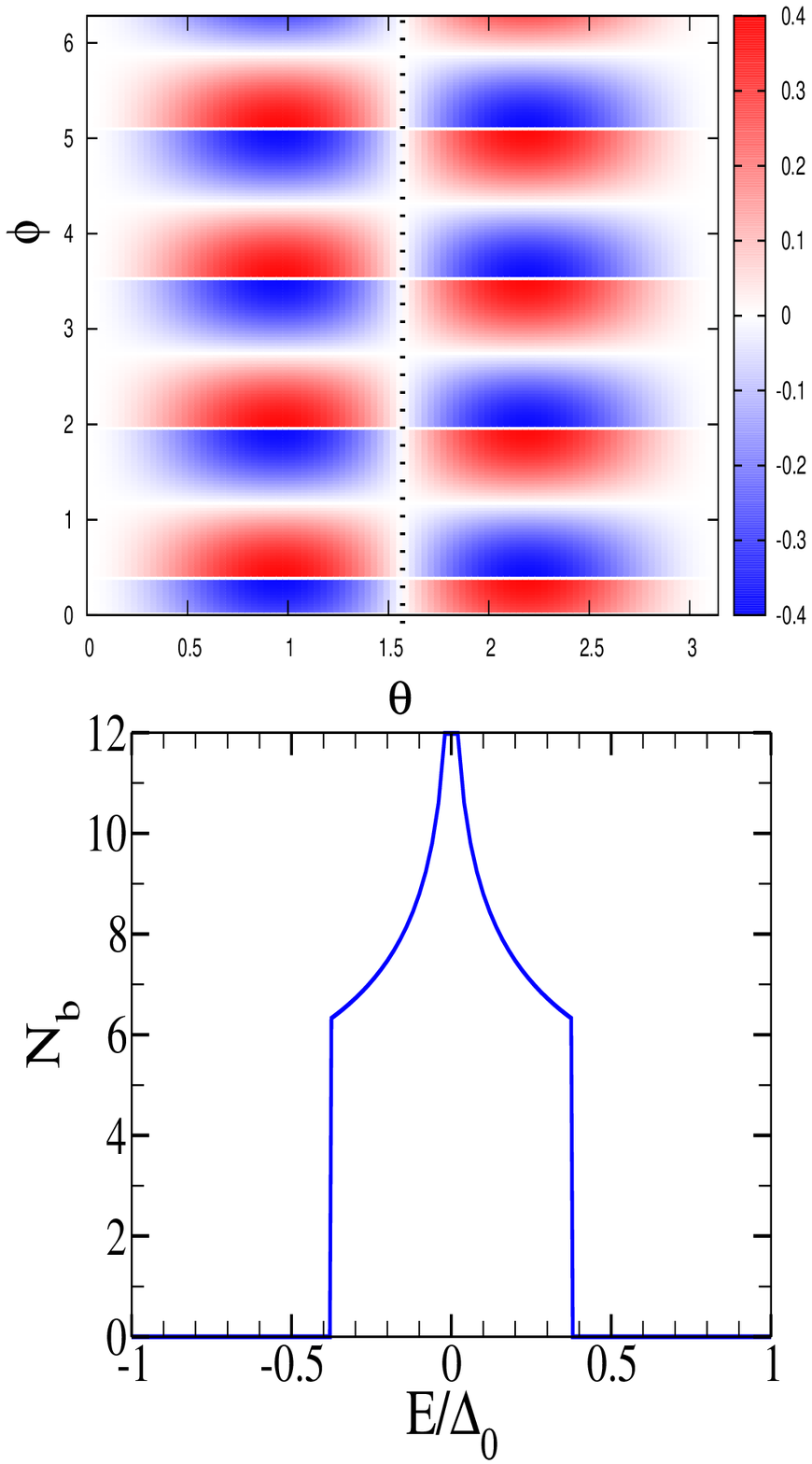}
\caption{(Color online) The ABS energy as a function of the direction of semiclassical propagation for $\chi=\pi/2$ (upper panel) and the corresponding DOS (lower panel), in the case of the chiral $f$-wave state ($E_{2u}$)
and $\bm{\hat{n}}||\bm{\hat{z}}$. The vertical dotted line at $\theta=\pi/2$ corresponds to the grazing trajectory (for which $v_{F,z}=0$).}
\label{fig:E2uC-nz}
\end{figure}

\underline{$\bm{\hat{n}}||\bm{\hat{z}}$.} It follows from Eq. (\ref{Ez_bound}) that
\begin{equation}
  E_b(\theta,\phi)=\Delta_{0}\sin^2\theta\cos\theta\cos\left(2\phi-\frac{\chi}{2}\right)\,\mathrm{sgn}\lb\sin\left(2\phi-\frac{\chi}{2}\right)\rb,
\end{equation}
which is discontinuous at $2\phi-\chi/2=0$ and $\pi$, see Appendix \ref{app: ABS-properties}. The ABS dispersion as a function of $\phi$ shifts upon changing $\chi$, therefore, 
the DOS does not depend on $\chi$ and we obtain from Eq. (\ref{DOSz}):
\be
  N_b(E)=\frac{2N_Fv_F}{\Delta_{0}}\int_{0}^{1} \frac{x\,dx}{\sqrt{x^2(1-x^2)^2-(E/\Delta_0)^2}}\Theta\left[x(1-x^2)-\frac{E}{\Delta_0}\right],
\ee
where $x=\cos\theta$.
The integral here is calculated numerically. The ABS energy for $\chi=\pi/2$ and the DOS for any $\chi$ are shown in Fig. \ref{fig:E2uC-nz}. The logarithmic singularity in the DOS at $E\to 0$ comes from the saddle points in the
ABS dispersion at $\theta=0,\pi$, \textit{i.e.}, for $\hat{\bk}_F$ perpendicular to the DW.

\subsection{$E_{2g}$ representation}

Finally, we consider the chiral $d$-wave state corresponding to the IREP $E_{2g}$, in which case $\Delta_+=\Delta_0e^{i\chi}(\hat k^2_{F,x}-\hat k^2_{F,y}-2i\hat k_{F,x}\hat k_{F,y})$ and 
$\Delta_-=\Delta_0(\hat k^2_{F,x}-\hat k^2_{F,y}+2i\hat k_{F,x}\hat k_{F,y})$. Therefore,
\be
\left.\begin{array}{l}
      \Delta_{+}(\theta, \phi)=\Delta_0 \sin^2\theta\, e^{i(\chi-2\phi)},\\ 
      \Delta_{-}(\theta, \phi)=\Delta_0 \sin^2\theta\, e^{2i\phi},
      \end{array}\right.
\ee
and $\beta=\tan(2\phi-\chi/2)$. 

\begin{figure}[H]
\vspace{-0.3in}
\includegraphics[width=5in,height=4in, angle=0]{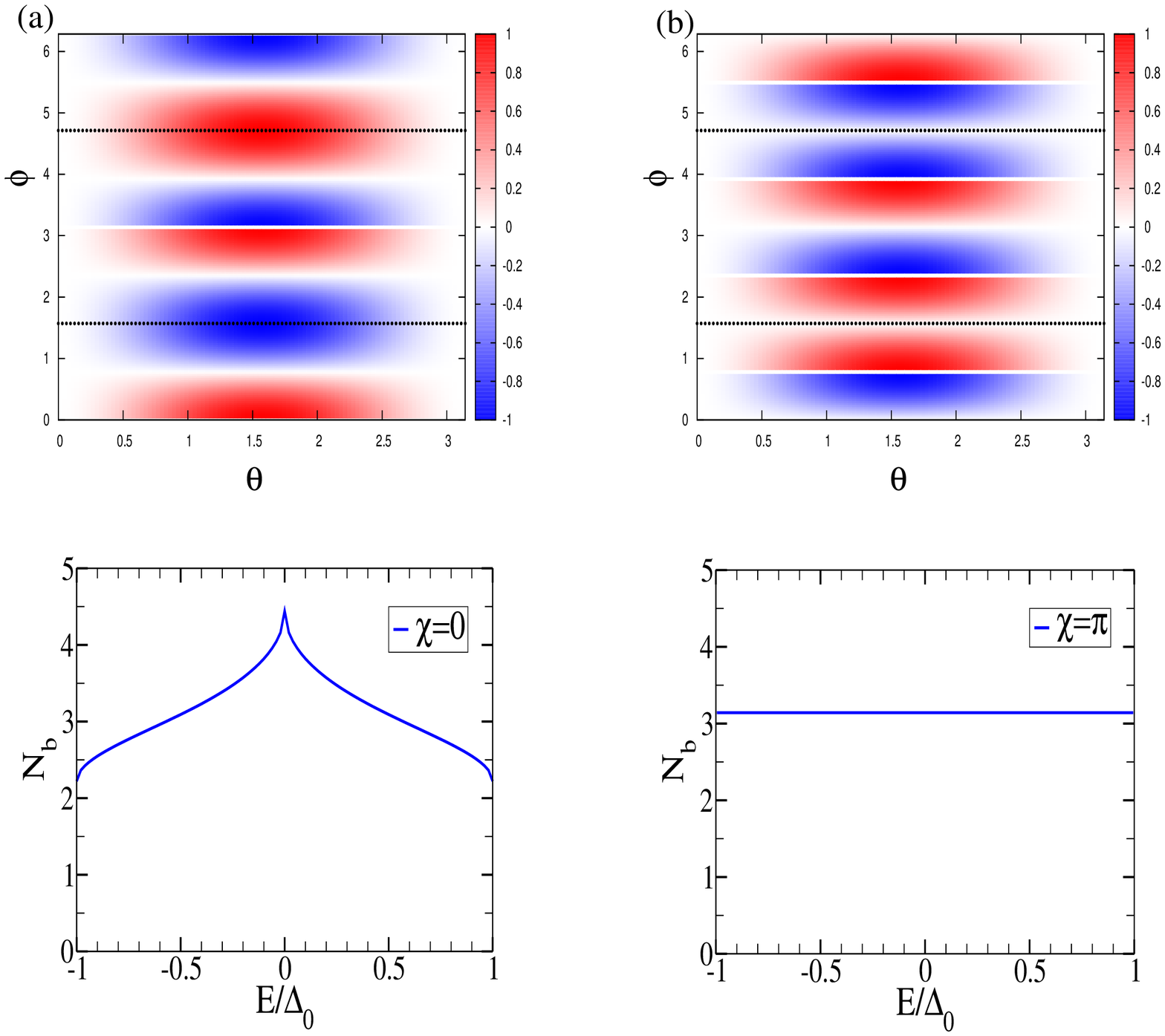}
\caption{The ABS energy as a function of the direction of semiclassical propagation (upper panels) and the corresponding DOS (lower panels), 
for $\chi=0$ and $\chi=\pi$, in the case of the chiral $d$-wave state ($E_{2g}$) and $\bm{\hat{n}}||\bm{\hat{x}}$. The grazing trajectories (for which $v_{F,x}=0$)
are shown at $\phi=\pi/2$ and $\phi=3\pi/2$ by horizontal dotted lines.}
\label{fig:E2gC-nx}
\end{figure}

\underline{$\bm{\hat{n}}||\bm{\hat{x}}$.} In this case, Eq. (\ref{Ex_bound}) takes the following form:
\begin{equation}
\label{Eb-E2g-chi}
  E_b(\theta,\phi)=\Delta_{0}\sin^2\theta\cos\left(2\phi-\frac{\chi}{2}\right)\,\mathrm{sgn}\lb\sin\left(2\phi-\frac{\chi}{2}\right)\cos\phi\rb,
\end{equation}
which is discontinuous at $2\phi-\chi/2=0,\pi$ and also at $\phi=\pi/2,3\pi/2$, see Appendix \ref{app: ABS-properties}. 
It has four lines of zeros at $\phi=(\chi\pm\pi)/4$ and $\phi=(\chi\pm 3\pi)/4$. The isolated second-order point zeros at $\theta=0$ and $\pi$ correspond to the trajectories parallel to the DW,
for which the Andreev approximation is not applicable. 
The ABS energy for $\chi=0$ and $\pi$ is shown in the upper panels of Fig. \ref{fig:E2gC-nx}. 

The DOS, see Eq. (\ref{DOSx}), can be calculated analytically for $\chi=0$ and $\chi=\pi$. Following the same steps as in the previous subsections, we obtain:
\be
N_b(E) = \frac{N_Fv_F}{\sqrt{2}\Delta_{0}} \lp \frac{\pi}{2} + \arcsin\frac{\sqrt{1-E/\Delta_{0}}}{\sqrt{1+E/\Delta_{0}}}\rp
\ee
for $\chi=0$, and 
\be
  N_b(E)=\frac{\pi N_Fv_F}{2\Delta_{0}}
\ee
for $\chi=\pi$, see Fig. \ref{fig:E2gC-nx}.

\underline{$\bm{\hat{n}}||\bm{\hat{z}}$.} We obtain from Eq. (\ref{Ez_bound}):
\begin{equation}
  E_b(\theta,\phi)=\Delta_{0}\sin^2\theta\cos\left(2\phi-\frac{\chi}{2}\right)\,\mathrm{sgn}\lb\sin\left(2\phi-\frac{\chi}{2}\right)\cos\theta\rb.
\end{equation}
The discontinuities of the ABS energy are located at $2\phi-\chi/2=0,\pi$ and also at $\theta=\pi/2$, see Appendix \ref{app: ABS-properties}. 
As in the previous subsections, the ABS dispersion as a function of $\phi$ shifts upon changing $\chi$, the DOS does not depend on $\chi$, and we obtain:
\be
N_b(E)=\frac{N_Fv_F}{\Delta_{0}}\int_{E/\Delta_0}^1\frac{dx}{\sqrt{x^2-(E/\Delta_0)^2}}=\frac{N_Fv_F}{\Delta_{0}} \ln\frac{\sqrt{1-(E/\Delta_0)^2}+1}{E/\Delta_0},
\ee
where $x=\sin^2\theta$. The ABS energy for $\chi=\pi/2$ and the DOS for any $\chi$ are shown in Fig. \ref{fig:E1gC-nz}. The logarithmic singularity in the DOS at $E\to 0$ comes from the saddle points in the
ABS dispersion at $\theta=0,\pi$, \textit{i.e.}, for $\hat{\bk}_F$ perpendicular to the DW.

\begin{figure}[H]
\vspace{-0.3in}
\includegraphics[width=2.5in,height=4in, angle=0]{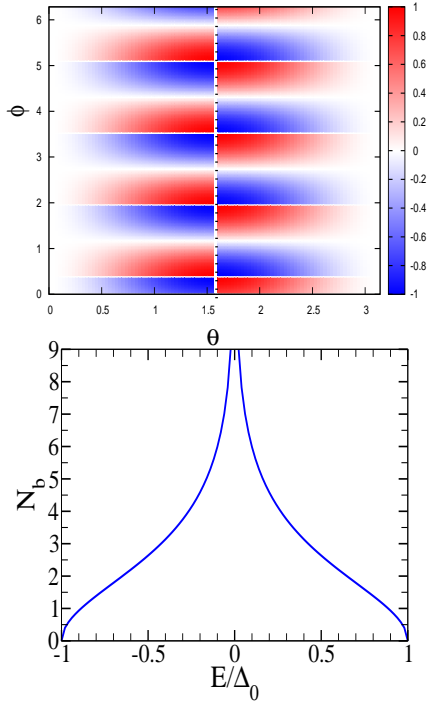}
\caption{(Color online) The ABS energy as a function of the direction of semiclassical propagation for $\chi=\pi/2$ (upper panel) and the corresponding DOS (lower panel), in the case of the chiral $d$-wave state ($E_{2g}$)
and $\bm{\hat{n}}||\bm{\hat{z}}$. The vertical dotted line at $\theta=\pi/2$ corresponds to the grazing trajectory (for which $v_{F,z}=0$).}
\label{fig:E2gC-nz}
\end{figure}

\subsection{Topological origin of the ABS zero modes}
\label{sec: topology}

The number of zero-energy ABS localized at the DW separating degenerate chiral states is determined by the difference between topological invariants characterizing the 
superconducting states in the bulk of the domains, which is known as the bulk-boundary correspondence.\cite{Volovik} As an illustration of this statement, we focus on the case of $\bm{\hat{n}}||\bm{\hat{x}}$.
To define the appropriate topological invariant, we introduce the Matsubara-like Green's function of the Bogoliubov quasiparticles in the bulk:
\begin{equation}
\label{BdG-GF}
  \hat G^{-1}(k_0,\bk)=ik_0-\hat H(\bk),
\end{equation}
where $ik_0$ is imaginary ``frequency'', $\bk=(k_x,k_y,k_z)$ takes values in the 3D Brillouin zone, and 
$$
  \hat H(\bk)=\lp\begin{array}{cc} \xi(\bk) & \Delta(\bk)\\ \Delta^*(\bk) & -\xi(\bk)\end{array}\rp
$$
is the Bogoliubov-de Gennes (BdG) Hamiltonian, with the gap function $\Delta(\bk)$. 
Since we consider only the singlet pairing and the triplet pairing with $\bm{d}\parallel\hat{\bm{z}}$, see Sec. \ref{sec:II}, the spin channels are decoupled and the $4\times 4$ BdG equations  
are reduced to a two-component (electron-hole, or Nambu) form for each spin. The BdG Hamiltonian can be written in the form $\hat H(\bk)=\bm{\nu}(\bk)\hat{\bm{\tau}}$, where $\hat{\bm{\tau}}$ are the Pauli matrices in the Nambu space and
$$
  \bm{\nu}(\bk)=\left(\begin{array}{c}
                         \re\Delta(\bk)\\
                         -\im\Delta(\bk)\\
                         \xi(\bk)
                        \end{array}\right).
$$
The eigenvalues of $\hat H(\bk)$ are given by $\pm E(\bk)$, where $E(\bk)=\sqrt{\xi^2(\bk)+|\Delta(\bk)|^2}$ the energy of the Bogoliubov fermionic excitations.

At given $k_z$, regarded as a parameter, one can define the following topological invariant:\cite{Volovik}
\begin{equation}
\label{N-kz-def}
  N(k_z)=-\frac{1}{24\pi^2}\int\mathrm{tr}(\hat Gd\hat G^{-1})^3.
\end{equation}
Here ``tr'' stands for the Nambu matrix trace, the powers of the 1-form $\hat Gd\hat G^{-1}$ should be understood in the sense of combined exterior and matrix multiplication, 
and the integration is performed over $k_0$ and $\bk_\perp=(k_x,k_y)$, with $\bk_\perp$ taking values in the 2D cross-section of the Brillouin zone by the constant $k_z$ plane. 
After some algebra, we obtain:
\begin{equation}
\label{N-kz-nu}
  N(k_z)=\frac{1}{4\pi^2}\int\frac{\bm{\nu}(d\bm{\nu}\times d\bm{\nu})dk_0}{(k_0^2+E^2)^2}=\frac{1}{8\pi}\int_{k_z=\mathrm{const}}\hat{\bm{\nu}}(d\hat{\bm{\nu}}\times d\hat{\bm{\nu}}),
\end{equation}
where $\hat{\bm{\nu}}=\bm{\nu}/|\bm{\nu}|$. We assume that the superconducting pairing is BCS-like and effective only near the Fermi surface. At given $k_z$, this results in the gap function being nonzero only near the Fermi line 
$\mathrm{FL}(k_z)$, which is the intersection of the Fermi surface and the constant $k_z$ plane. We represent the gap function in the form $\Delta(\bk)=|\Delta(\bk)|e^{i\varphi(\bk)}$, and assume
that there are no gap nodes, i.e. the gap magnitude does not vanish anywhere on the Fermi line. Then it follows from Eq. (\ref{N-kz-nu}) that 
\begin{equation}
\label{N-kz-final}
  N(k_z)=\frac{1}{2\pi}\oint_{\mathrm{FL}(k_z)}d\varphi=\left.\frac{\Delta\varphi}{2\pi}\right|_{\mathrm{FL}(k_z)},
\end{equation}
therefore the topological invariant (\ref{N-kz-def}) is nothing but the phase winding number of the gap function around the cross-section of the Fermi surface at given $k_z$. Assuming a spherical Fermi surface, the cross-section is a circle of 
radius $k_{F,\perp}=\sqrt{k_F^2-k_z^2}$.
For the superconducting states considered above, the topological invariant (\ref{N-kz-final}) takes opposite nonzero values for the states of opposite chirality, see Table \ref{tab:N-kz}. The topological invariants are not 
defined at the bulk gap nodes, i.e. at $k_z=\pm k_F$ for all four IREPs and additionally at $k_z=0$ for the IREPs $E_{1g}$ and $E_{2u}$. 

According to Eqs. (\ref{Eb-E1u-chi}), (\ref{Eb-E1g-chi}), (\ref{Eb-E2u-chi}), and (\ref{Eb-E2g-chi}), the ABS energy for all four IREPs can be written in the following form:
\begin{equation}
\label{Eb-common}
  E_b(\theta,\phi)=\Delta_{0}f_\Gamma(\theta)\cos\left(n\phi-\frac{\chi}{2}\right)\,\mathrm{sgn}\lb\sin\left(n\phi-\frac{\chi}{2}\right)\cos\phi\rb,
\end{equation}
where $n=1$ for $\Gamma=E_{1u},E_{1g}$ and $n=2$ for $\Gamma=E_{2u},E_{2g}$, and the function $f_\Gamma(\theta)$ depends on the IREP. At fixed $k_z=k_F\cos\theta$, the last expression vanishes at some values of $\phi$, corresponding to 
the ABS zero modes. It is easy to see that there are $2n$ zero modes: for $\Gamma=E_{1u},E_{1g}$ they correspond to $\phi=(\chi\pm\pi)/2$, while for $\Gamma=E_{2u},E_{2g}$ they correspond to 
$\phi=(\chi\pm\pi)/4$ and $\phi=(\chi\pm 3\pi)/4$. 

One can define the algebraic number $\nu(k_z)$ of the ABS zero modes as the number of positive-velocity modes minus the number of negative-velocity modes. According to the bulk-boundary correspondence (Ref. \onlinecite{Volovik}), 
$\nu$ is equal to the difference between the topological invariants in the bulk of the two domains:
\begin{equation}
\label{bulk-boundary}
  \nu(k_z)=N(k_z)|_{x>0}-N(k_z)|_{x<0}.
\end{equation}
Expressing the ABS energy, see Eq. (\ref{Eb-common}), in terms of $k_y=k_{F,\perp}\sin\phi$, one can show that the ABS zero modes propagate along the DW in the same direction:
$\mathrm{sgn}(\partial E_b/\partial k_y)=\mathrm{sgn}(\partial E_b/\partial\phi)\,\mathrm{sgn}(\cos\phi)=-1$, therefore $\nu=-2n$. 
On the other hand, it follows from Table \ref{tab:N-kz} that $N(k_z)|_{x>0}=-n$ and $N(k_z)|_{x<0}=n$, which means that Eq. (\ref{bulk-boundary}) is indeed satisfied. Taking into account the doubling of the degrees 
of freedom due to spin, the total number of the ABS zero modes localized near the DW is equal to $4n$, at given $k_z$. Note that the same topological argument can be used to 
prove the existence of zero-energy ABS near the surface of an unconventional superconductor, see Ref. \onlinecite{ZBCP}. For UPt$_3$, it was done recently in Ref. \onlinecite{GN15}.

\begin{table}
\caption{Topological invariant, Eq. (\ref{N-kz-final}), for the chiral states corresponding to the 2D IREPs of $D_{6h}$.}
\begin{tabular}{|c|c|c|}
\hline 
   IREP   & gap function & $N(k_z)$ \\
\hline 
$E_{1u}$  & $k_x\pm ik_y$ & $\pm 1$ \\
\hline 
$E_{1g}$  & $k_{z}(k_x \pm i k_y)$ & $\pm 1$ \\
\hline 
$E_{2u}$ & $k_z(k^{2}_x-k^{2}_y\pm 2ik_xk_y)$ & $\pm 2$ \\
\hline 
$E_{2g}$ & $k^{2}_x-k^{2}_y\pm 2ik_xk_y$ & $\pm 2$ \\
\hline 
\end{tabular}
\label{tab:N-kz} 
\end{table}

\section{Conclusion}
\label{sec:V}

We have found that the DWs separating degenerate TRS-breaking superconducting states in a 3D hexagonal crystal always create the quasiparticle ABS, for all directions of the semiclassical propagation.
We have considered all four 2D IREPs of the point group $D_{6h}$ (two singlet and two triplet cases) and two orientations of the DW, parallel and perpendicular to the $z$ axis. 
The ABS spectrum strongly depends on the order parameter symmetry and the DW orientation. Additionally, it is affected by the Josephson phase difference $\chi$ between the
domains, which is determined by the microscopic parameters. If the DW is parallel to the $z$ axis, then there is a significant difference between 
the chiral states $(1,\pm i)$ (corresponding to $\chi=0$) and $(\pm 1,i)$ (corresponding to $\chi=\pi$), which can be treated analytically. 
The spectrum of the DW ABS's can be probed in tunneling experiments by measuring their DOS, which has very different energy dependence from that of the bulk quasiparticles. 
We have calculated the DOS per unit area of the DW and found a widely varying behaviour, the most prominent feature being the logarithmic van Hove singularity at zero energy, which is present 
in several cases. 

Despite the qualitative sensitivity of the DOS to the microscopic parameters that cannot be easily controlled in experiment, we can still make some firm predictions for the DW effects on the tunneling measurements in UPt$_3$.
First, there is strong evidence that the gap symmetry in the $B$ phase of UPt$_3$ is described by the chiral $f$-wave state corresponding to the IREP $E_{2u}$. If this is the case, then our results in Sec. \ref{sec: E-2u}
indicate that the zero-energy singularity in the DOS is a universal feature, which, in contrast to the other three IREPs, is present for both orientations of the DW and for all values of $\chi$. 
Second, if the DW is perpendicular to the $z$ axis, then the DOS does not actually depend on $\chi$, showing different behaviour for the four IREPs: a broad dome-like maximum for $E_{1u}$, a constant for $E_{1g}$, the zero-energy singularity with
two sharp edges for $E_{2u}$, and the zero-energy singularity without sharp edges for $E_{2g}$. We hope that these features can be directly probed in tunneling experiments, thus shedding light on the presence of the DWs as well as
the underlying pairing symmetry in UPt$_3$ and other hexagonal superconductors.

\acknowledgments

This work was supported by a Discovery Grant from the Natural Sciences and Engineering Research Council of Canada.

\appendix

\section{Discontinuities of the ABS spectrum}
\label{app: ABS-properties}

In all cases studied in this paper the gap function has the same magnitude on both sides of the DW: $\Delta_\pm=\Delta e^{i\varphi_\pm}$. 
Therefore, $\gamma=\Delta_-/\Delta_+=e^{i\Phi}$, where $\Phi=\varphi_--\varphi_+$. We obtain from Eqs. (\ref{beta-def}) and (\ref{E_bound}): 
$$
  \beta=\tan\left(\frac{\Phi}{2}\right)
$$
and 
\begin{equation}
\label{app-E_b}
  E_b(\hat{\bk}_F)=\Delta\cos\left(\frac{\Phi}{2}\right)\mathrm{sgn}\left[\sin\left(\frac{\Phi}{2}\right)(\hat{\bm{v}}_{F}\cdot\hat{\bm{n}})\right].
\end{equation}
Both $\Delta$ and $\Phi$ depend on the direction of semiclassical propagation, characterized by the Fermi-surface wavevector $\bk_F$. At each $\bk_F$, we have $|E_b|\leq\Delta$, therefore the ABS are disconnected from the bulk states.

It follows from Eq. (\ref{app-E_b}) that the ABS energy is not defined for some $\hat{\bk}_F$. 
It is discontinuous at $\Phi=0\mod 2\pi$, which corresponds to $\Delta_+=\Delta_-$. For such semiclassical trajectories, the DW is ``invisible'' to the quasiparticles. The ABS energy is also discontinuous at $\hat{\bm{v}}_{F}\perp\hat{\bm{n}}$, 
i.e. when the quasiparticles move parallel to the DW. For such trajectories, the Andreev approximation itself is not applicable.

\section{Quasiparticle DOS near a domain wall}
\label{app: DOS-derivation}

The local quasiparticle DOS for both spin projections is given by the following expression:
$$
  N(\br,E)=-\frac{1}{\pi}\sum_{\alpha=\uparrow,\downarrow}\im G^R_{\alpha\alpha}(\br,\br;E).
$$
Here $G^R$ is the retarded Green's function, which is obtained in the standard fashion, by analytically continuing the Fourier transform of the Matsubara Green's function
$G_{\alpha}(\br_1,\br_2;\tau)=-\langle T_\tau\psi_\alpha(\br_1,\tau)\psi^\dagger_\beta(\br_2,0)\rangle$ to real frequencies.\cite{AGD}
Next, we represent the field operators as
\begin{eqnarray*}
    \psi_\alpha(\br)=\sum_a[u_a(\br,\alpha)\gamma_a+v_a^*(\br,\alpha)\gamma_a^\dagger],\\
    \psi^\dagger_\alpha(\br)=\sum_a[v_a(\br,\alpha)\gamma_a+u_a^*(\br,\alpha)\gamma_a^\dagger],
\end{eqnarray*}
where $\gamma^\dagger_a,\gamma_a$ are the creation and annihilation operators of the Bogoliubov quasiparticles and the quantum numbers $a$ label the upper half of the spectrum ($E_a\geq 0$) of the $4\times 4$ Bogoliubov-de Gennes Hamiltonian.

Both in the singlet case and in the triplet case with $\bm{d}\parallel\hat{\bm{z}}$, the spin channels decouple and the local DOS becomes
\begin{equation}
\label{app: DOS-2by2}
  N(\br,E)=2\sum_a\left[|u_a(\br)|^2\delta(E-E_a)+|v_a(\br)|^2\delta(E+E_a)\right],
\end{equation}
where the two-component Nambu spinor satisfies the following equation:
\begin{equation}
\label{app: BdG-2by2}
  \left(\begin{array}{cc}
 	\hat\xi & \hat\Delta\\
	\hat\Delta^\dagger & -\hat\xi
	\end{array}\right)
   \left(\begin{array}{c}
 	u_a\\
	v_a
	\end{array}\right)=E_a
   \left(\begin{array}{c}
 	u_a\\
	v_a
	\end{array}\right).
\end{equation}
Due to the electron-hole symmetry of the BdG spectrum, one can focus only on the electron-like branch with $E\geq 0$. 

If the order parameter depends only on $x$, then the normalized solutions of Eq. (\ref{app: BdG-2by2}) have the form 
\begin{eqnarray*}
  && \left(\begin{array}{c}
 	u(\br)\\
	v(\br)
	\end{array}\right)=\frac{1}{\sqrt{L_yL_z}}
	\left(\begin{array}{c}
 	u(x)\\
	v(x)
	\end{array}\right)e^{ik_yy}e^{ik_zz},\\
  && \int_{-\infty}^\infty dx\;(|u|^2+|v|^2)=1.
\end{eqnarray*}
One can define the quasiparticle DOS per unit area in the $yz$ plane as follows:
\begin{equation}
\label{app: N-area}
  N(E)=\int_{-\infty}^\infty dx\; N(\br,E).
\end{equation}
For the ABS, we have $|u|^2=|v|^2$, therefore $\int_{-\infty}^\infty|u|^2dx=1/2$. Inserting this last expression in Eqs. (\ref{app: N-area}) and (\ref{app: DOS-2by2}), we arrive at Eq. (\ref{DOS-x-general}).

\end{document}